\documentclass[journal]{IEEEtran}
\usepackage{tabularx}
\usepackage{multirow}
\usepackage{subcaption}
\usepackage{ragged2e}
\usepackage{url}

\ifCLASSINFOpdf
\usepackage[pdftex]{graphicx}
\else
\fi
\ifCLASSOPTIONcompsoc
    \usepackage[caption=false,font=normalsize,labelfont=sf,textfont=sf]{subfig}
\else
    \usepackage[caption=false,font=footnotesize]{subfig}
\fi
\begin{document}
%

\title{Policy Design in Zero-Trust Distributed Networks: Challenges and Solutions}
%
%

\author{Fannya R. Sandjaja, Ayesha A. Majeed, Abdullah Abdullah, Gyan Wickremasinghe, \\Karen Rafferty and Vishal Sharma* 
        \thanks{F. R. Sandjaja,  A. A. Majeed, G. Wickremasinghe, K. Rafferty and V. Sharma are with the School of Electronics, Electrical Engineering and Computer Science (EEECS), Queen's University Belfast QUB), Email: \{fsandjaja, ayesha.abdulmajeed, gwickremasinghe01, k.rafferty, v.sharma\}@qub.ac.uk.
        A. Abdullah is with Sir William Dunn School, Oxford University, Email: abdullah.abdullah@path.ox.ac.uk}

\thanks{Manuscript received ...}}

%
%


\markboth{Preprint}%
{Shell \MakeLowercase{\textit{et al.}}: Bare Demo of IEEEtran.cls for IEEE Journals}
%



\maketitle

\begin{abstract} 
Traditional security architectures are becoming more vulnerable to distributed attacks due to significant dependence on trust. This will further escalate when implementing agentic AI within the systems, as more components must be secured over a similar distributed space. These scenarios can be observed in consumer technologies, such as the dense Internet of things (IoT). Here, zero-trust architecture (ZTA) can be seen as a potential solution, which relies on a key principle of not giving users explicit trust, instead always verifying their privileges whenever a request is made. However, the overall security in ZTA is managed through its policies, and unverified policies can lead to unauthorized access. Thus, this paper explores challenges and solutions for ZTA policy design in the context of distributed networks, which is referred to as zero-trust distributed networks (ZTDN). This is followed by a case-study on formal verification of policies using UPPAAL. Subsequently, the importance of accountability and responsibility in the system's security is discussed.
\end{abstract}

\begin{IEEEkeywords}
 Zero-Trust, Agentic AI, Policy design, Security, Formal verification
\end{IEEEkeywords}

%
\IEEEpeerreviewmaketitle

\section{Introduction}

\begin{figure*}[h]
\centerline{\includegraphics[width=30pc, trim=10 360 10 70, clip]{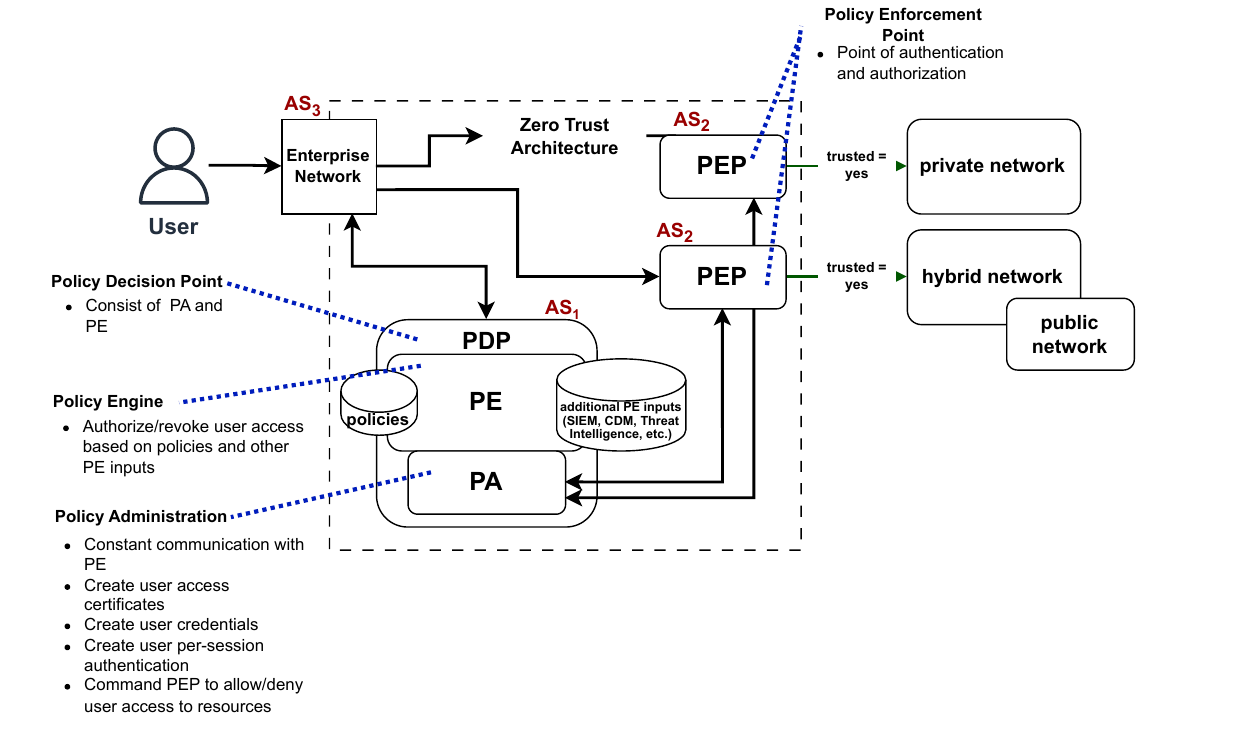}}
\caption{An exemplary illustration of stand-alone zero-trust distributed network (ZTDN)  ($AS_x$: Attack Surface x).}
\label{figure1}
\end{figure*}

Security is undoubtedly one of the most critical issues in consumer technology seen from the sharp rise in attacks against enterprises. Puthal et al. (2017) \cite{puthal2017} have mentioned that traditional perimeter-based security can no longer sustain the improvement of cyber attacks in recent years. Hence, more novel solutions are needed to defend against new types of cyber security threats, such as those generated by data breaches \cite{poirrier2025}. A security model, namely the zero-trust architecture (ZTA), proposed in 2010 by Kindervag et al. \cite{kindervag2010}, has been gradually gaining more traction lately because it can help solve most of the current security scenarios, particularly in the context of distributed networks, which are explored in this paper as zero-trust distributed networks (ZTDN). Zero-trust (ZT) operates based on the \textit{trust nothing, verify everything} concept, which assumes that nothing in the system is inherently trustworthy \cite{puthal2017}, \cite{dhruva2024}. The security model in \cite{kindervag2010} adopts a few principles, namely least privilege, trust no one, micro-segmentation, and always verify. Applying this concept to distributed networks can help strengthen the defense against vulnerabilities. Distributed networks are widely used for migrating services to the cloud and creating separate data centers. This raises security concerns, such as malicious attackers gaining access to all resources through a loophole from a single-point entry that can be addressed with ZT, for example, through its principle of micro-segmentation \cite{lee2025}. Figure \ref{figure1} displays an example of a use case for implementing ZT in distributed networks.

The components of ZTA include the policy enforcement point (PEP) and the policy decision point (PDP), which consist of the policy engine (PE) and policy administration (PA). A PEP is defined as a system that oversees every connection request that comes through the network. It communicates with the PA to forward requests from the user and receive updates on whether the user is trusted based on the policies \cite{fernandez2024}. Here, the PA can be trusted, and the summary of individual components is provided in Figure \ref{figure1}.

In the given Figure \ref{figure1}, a user is someone who wants to gain access to enterprise systems, whether that is an enterprise's private or hybrid network. System grants trust to the user to allow entry. For this purpose, security policies must be in place to protect the network and be selective of who gets admission to the resources that are being requested, and in ZTDN, policies play a vital role in allowing and denying access to a system's resources.

For each network and its resources, a PEP is in place as a point of authentication and authorization. This is done even though the user has been authenticated and authorized when initiating access to the enterprise network via user login credentials or other methods. This is a part of the ZTA principle of continuous monitoring and constantly verifying the actions of the user.

Jung et al. (2022) \cite{jung2022} have explained that a PA acts as an administrator for the network connection. This includes communicating with the PE on whether the policy is recognized and whether the user identity is authorized to access the resources that have been requested. PA can create user access certificates, credentials, and per-session authentication. When a user is trusted and their access is authorized, the PA commands the PEP to open the communication channel for the user to the resources. However, if the user is untrustworthy and their credentials are unauthorized, the PA will instruct the PEP to block all communication channels for the user. PE also has the capability to revoke user access using supplemented policies, and it records the history of access requests, which can help in making future decisions \cite{jung2022}.

Even with the claims of being able to defend against the developments of attacks, there are still some challenges that can be faced while utilizing ZT in distributed networks, such as setting suitable access policies for the authorized user or insider threats that have initial access to the system \cite{tsai2024}. 

Most recently, Cisco has published its ZT network security architecture, which mainly focuses on ZT security in its distributed networks-related products. Other applications include the implementation of ZTA in cloud networks by Netskope and internal infrastructure networks initiated by Google's BeyondCorp \cite{feng2023}.

Rais et al. (2024) \cite{rais2024} have explained that there has been no standardization for ZT policies, and industry-wide standards to define policies are still an ongoing development. It is understandable that policies in ZT are distinguishable mainly because they focus more on the logical components of the network. Further research on policies specifically in ZTDN is necessary to tackle the ever-evolving threats in this domain. 

This article discusses the current development of ZTA, specifically in the context of distributed networks (ZTDN). It explores the challenges of ZT in distributed networks and discusses existing solutions to these challenges. This article also highlights the importance of responsible policy for ZTDN. Finally, an exemplary case study of formally verifying policies in ZTDN is presented using UPPAAL, an open-source tool that act as a protocol verifier and can be applied to verify and validate policies \cite{uppaal}, followed by future research directions.

\section{Challenges and Solutions in ZTA in the Context of ZTDN}

\begin{table*}[!ht]
    \renewcommand{\arraystretch}{1.5}
    \caption{Key challenges in ZTDN ($AS_{x}$: Attack Surface x).}
    \centering
    \begin{tabularx}{\linewidth}{| X | X | X | X | X | X | X |}
    \hline
        \multirow{2}*{Area of Impact} & 
            \multicolumn{1}{c|}{\multirow{2}{*}{Key Challenges}} &
            \multicolumn{1}{c|}{\multirow{2}{*}{Potential Attacks}} &
            \multicolumn{2}{c|}{Rule-based ZTDN} &
            \multicolumn{2}{c|}{Agentic AI} \\ \cline{4-7} 
            &
            \multicolumn{1}{c|}{} &
            \multicolumn{1}{c|}{} &
            \multicolumn{1}{c|}{SLAs} &
            \centering KPIs &
            \multicolumn{1}{c|}{SLAs} &
            \centering\arraybackslash KPIs \\
    \hline
        \multirow{4}{20mm}{Policy Engine}
        & \RaggedRight No quantitative trust evaluation \cite{ge2024} & \RaggedRight Conflicting network access (Figure~\ref{figure1} ($AS_1$)) & \RaggedRight Policies for trust evaluation algorithm & Unauthorized attempts frequency analysis & \RaggedRight Procedure for AI forensics & \RaggedRight Activity log \\\cline{2-7}
        
        & \RaggedRight No thresholds for trust score \cite{bradatsch2023} & \RaggedRight Data manipulation (Figure~\ref{figure1} ($AS_1$)) & \RaggedRight Policies for enterprise trust score threshold & \RaggedRight Response time and Incident prevention ratio & \RaggedRight Procedure of default behavior & \RaggedRight Results analysis and Misbehavior detection \\\cline{2-7}
        
        & \RaggedRight Lack of access control rules \cite{huber2024} & \RaggedRight Brute-force attacks, Compromised access credentials, and Insider threats (Figure~\ref{figure1} ($AS_1$,$AS_3$)) & \RaggedRight Rules for enterprise access policy & \RaggedRight Device availability and Device inventory log & \RaggedRight Procedure of agents' identification & \RaggedRight Availability and Accessibility\\\cline{2-7}
        
        & \RaggedRight Lack of trust awareness in policy language \cite{dimitrakos2020} & \RaggedRight Malware, Social engineering attacks, and Phishing (Figure~\ref{figure1} ($AS_1$, $AS_3$)) & \RaggedRight Configurable and responsible policy design with defined attributes & \RaggedRight Regulatory requirements, Latency, and Breach attempt counts & \RaggedRight Procedure of interrupting/terminating process & Response time\\
        
    \hline 
        Policy Enforcement Point & \RaggedRight Component failure \cite{spanier2023} & \RaggedRight Brute-force attacks, DDoS (Figure~\ref{figure1} ($AS_2$, $AS_3$)) & \RaggedRight Component failure management and resolution & \RaggedRight Breach attempt counts and Response time & \RaggedRight Continuous monitoring of AI agents & \RaggedRight Response time, Availability, and Accessibility\\
    \hline
    \end{tabularx}
    \label{table_1}
\end{table*}

ZTA promises a better security structure than traditional security architecture, but there are some associated challenges when implementing it in distributed network scenarios. The primary challenge is the lack of research on the policy engine, which can lead to confusion about who is responsible for approving access and associated accountability. Earlier literature has stated some challenges that are faced when trying to implement the concept of ZT, which include the lack of methods for quantifying trust in users, the lack of defined thresholds for said trust, the lack of clearly defined rules for the access control, the lack of trust awareness in policy language, and scenarios where ZT component fails. 

Ge and Zhu (2024) \cite{ge2024} have focused on a few challenges when dealing with ZT in the context of a 5G Internet of things (IoT) network and have mentioned that one of the problems is the lack of quantitative definition and measurement of trust from the agent, which might impact how the policies are planned and designed. In their paper, agent refers to an entity in an agent-centric trust evaluation framework. Trust is a vital part of defining policies to ensure that access permission is only granted to continuously authenticated and authorized users. This means that a distinct responsible policy should be in place, in particularly dealing with the ZTDN scenarios. In their paper \cite{ge2024}, the authors have proposed a mathematical approach to quantify trust, exploring the possibility of utilizing game theory as part of the policy engine plan and design.

Further down the challenges in the trust aspect, Bradatsch et al. (2023) \cite{bradatsch2023} have discussed the gap in the available list of attributes that can be placed into the policy engine. Their work has focused on the trust algorithm and addressed the challenge of not having a clear threshold to be compared with the trust scores attained through the policy engine. Further, the authors have highlighted the need for novel solutions in trust algorithms and emphasized how access decisions can be made from specific actions \cite{bradatsch2023}. They have considered a novel solution for defining trust in policy engines that quantifies the attributes needed to allow access to enterprise resources by defining a new method for setting the threshold for trust scores. 

Another perspective on the challenges in ZTDN comes from a paper by Spanier et al. (2023) \cite{spanier2023}, who have mentioned that while ZTA can reduce the damage caused by malicious attacks in the networks, it is often observed that centralized authentication and PEP can be a hindrance that causes a single point of failure. Taking this into consideration in ZTDN, it is essential to be critical of the failure of components at the individual level. There is a lack of solutions for hosts to communicate with each other when a particular element fails in the network. Thus, the architecture must be resilient so that when one component fails, another component can still be available. However, it is unclear whether a backup component will be available in the system and if this will result in additional overheads or redundancy. The authors in \cite{spanier2023} have explored the prospect of applying blockchain authentication for user verification, which may be fed into a policy validator. In distributed networks, to ensure that all systems adhere to the same policy, a policy validator can be implemented to emphasize the responsibility of the policies in the engine. Their approach includes decentralizing the policy engine to make the system more efficient in decision-making.

From the access control and policy perspectives in ZTA, Huber and Kandah (2024) \cite{huber2024} have discussed some challenges related to maintaining access policy. They have emphasized the lack of rules in the ZTA's access control system. Malicious attacks can compromise the system due to existing vulnerabilities without a clear definition of who, what, when, and how a user can access the network. The authors in \cite{huber2024} have suggested a solution of integrating a trust management component into ZTA, expecting to increase the architecture's security posture. Their proposed architecture, Zero Trust+, focuses on dynamically authenticating and authorizing users in real-time with adjustments according to the user's behavior. This direction of research is further supported by the works of Dimitrakos et al. (2020) \cite{dimitrakos2020}, who have explored the possibility of further enhancing ZTA because of found weaknesses, namely, the lack of certification or service level agreements (SLAs). In their work, the authors have emphasized these weaknesses as a significant concern in unpredictable security or privacy issues. Their work further suggests that the current policy language does not include trust awareness as one of its main components, hence limiting the trust and access policy. Considering such aspects of strengthening ZTA, it is further advisable to explore policy agreements for scenarios that encounter unpredictable adversaries. Dimitrakos et al. (2020) \cite{dimitrakos2020} have recommended incorporating attribute-based access control with a trust level evaluation engine for a policy evaluation engine in ZTA. Their solution, in particular, focuses on consumer IoT. Their security posture combines dynamic authorization, usage control (UCON), and probabilistic trust assessment, which is referred to as UCON+, simultaneously supporting policy and trust level evaluation, attribute value retrieval, and policy parsing.

Most key challenges from earlier research show a correlation between the need for better policy design and cultivating trust in the user. These new solutions have been impactful in the field of ZTDN. However, the PE component, particularly the policies itself, has not yet been broadly explored. This is critical in ensuring the user is deemed trustworthy to access requested resources. This leads to the current issue of developing and implementing responsible policy design. With the increasing usage of artificial intelligence (AI) tools to design policies, responsible policy design ensures that all policies to secure enterprise networks comply with each specific country's regulations. This is where SLAs may come into play, with an explicit agreement between the network and users, as it can be helpful to provide transparent terms and conditions about the usage and how users can access resources, as the least privilege tenet is applied. From the enterprise perspective, setting up these SLAs would also help set the baseline on how the services use users' personal data. Additionally, continuously monitoring the system's key performance indicators (KPIs) can be beneficial as an alert if unexpected metrics show up. 

\begin{figure*}[!ht]
    \centerline{\includegraphics[width=25pc, trim=40 120 40 200, clip]{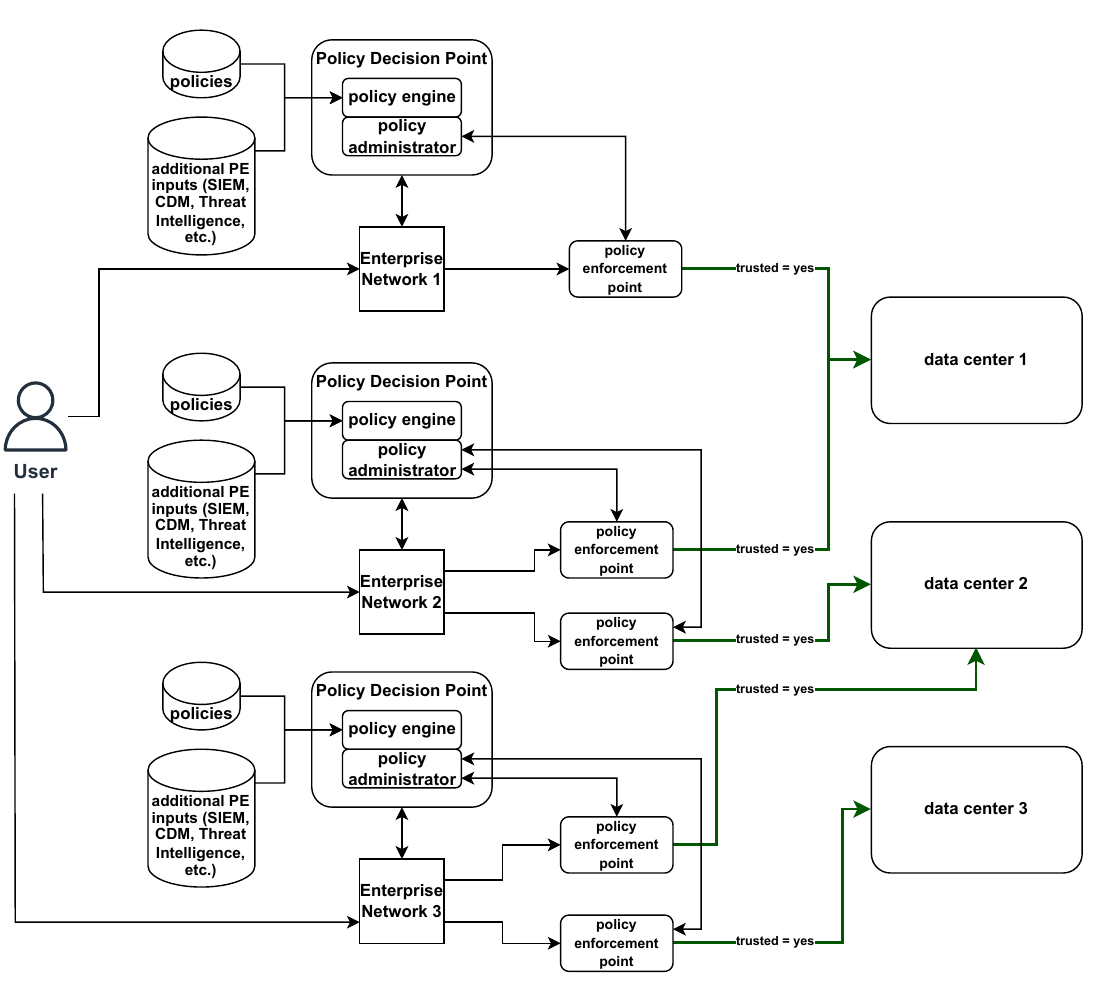}}
    \caption{An exemplary illustration of ZTDN architecture.}
    \label{figure2}
\end{figure*}
In this regard, setting up individual agents could be efficient and offer better control of systems design. However, the recent developments in agentic AI suggest that improper configurations can make the system more vulnerable. Here, SLAs that can be beneficial to be put in place include, SLAs for AI agent's default behavior, reliable tag identifier of each AI agent, a procedure to interrupt or terminate the process of agentic AI, consent of using AI agents, regulations of AI, what data protection law is being used for the agentic AI and which law enforcement documents does the agentic AI comply with.

On the system side, KPIs can be monitored continuously to safeguard the accessibility of the AI agents. These include the response time, availability, activity log, and result analysis of these agents in the system. This ensures that whenever the response time is longer or shorter than normal usage, that could be an indication of suspicious activity. This also applies to the availability and activity log of the agents. One of the KPIs could be result analysis that can provide a more in-depth observation of how the AI agents return a response to a user's query. Here, semantic analysis can be used if there might be sensitive information being exchanged between the user and agents, which could be an indication of a malicious act in the system. Table \ref{table_1} summarizes all critical challenges in rule-based and agentic AI driven ZTDN, the potential attacks that could occur, and provides insight into what type of SLAs can be implemented with the KPIs that could be monitored continuously.

\section{Challenges in Responsible Policy Design}
There are several problems when it comes to responsible policy design. Firstly, the methodology by which the system gives access to entities. Different ways have been explored to determine trust scores and levels, ranging from mathematical approaches to utilizing novel technology solutions like blockchain and game theory. But how are the standards defined for these attributes? How can different systems adjust their thresholds? These are incredibly challenging questions in distributed networks particularly where the trust threshold may not be the same for every entity. Hence, a responsible policy, possibly in the form of an SLA, is needed as enterprise networks are vastly different and include diverse requirements, structures, and compulsory policies \cite{ge2022}.

Secondly, the methodology by which the same user is treated differently across different network components, which is also provided as an exemplary illustration of ZTDN architecture in Figure \ref{figure2}. Here, in this scenario, a user tries to access three different enterprise networks, where the first enterprise network evaluates the user trust score to be above the threshold. In that case, the user is allowed access to resources. However, in enterprise network 2 and 3, the user trust scores are below the threshold, which means the user is untrusted. This can be a vulnerability in the system if the data centers are shared across these enterprise networks. It might make the system more prone to attacks, especially from within the networks. As such, a policy design solution that can be implemented industry-wide would be helpful for the future development of ZT in distributed networks \cite{davis2015}.

Lastly, there is an issue with accountability in ZT policies. There are no clear rules on who is responsible for the policies fed into the PE. This could also be related to network evidence gathering and, depending on the use of agentic AI, it may additionally require AI forensics. In this regard, who will be responsible if there is a policy design flaw? Such concerns need a clear structure of people overseeing decisions on creating, monitoring, and removing policies in ZTDN \cite{rose2020}. 

\begin{figure*}[h]
\centerline{\includegraphics[width=34pc, trim=0 300 0 0, clip]{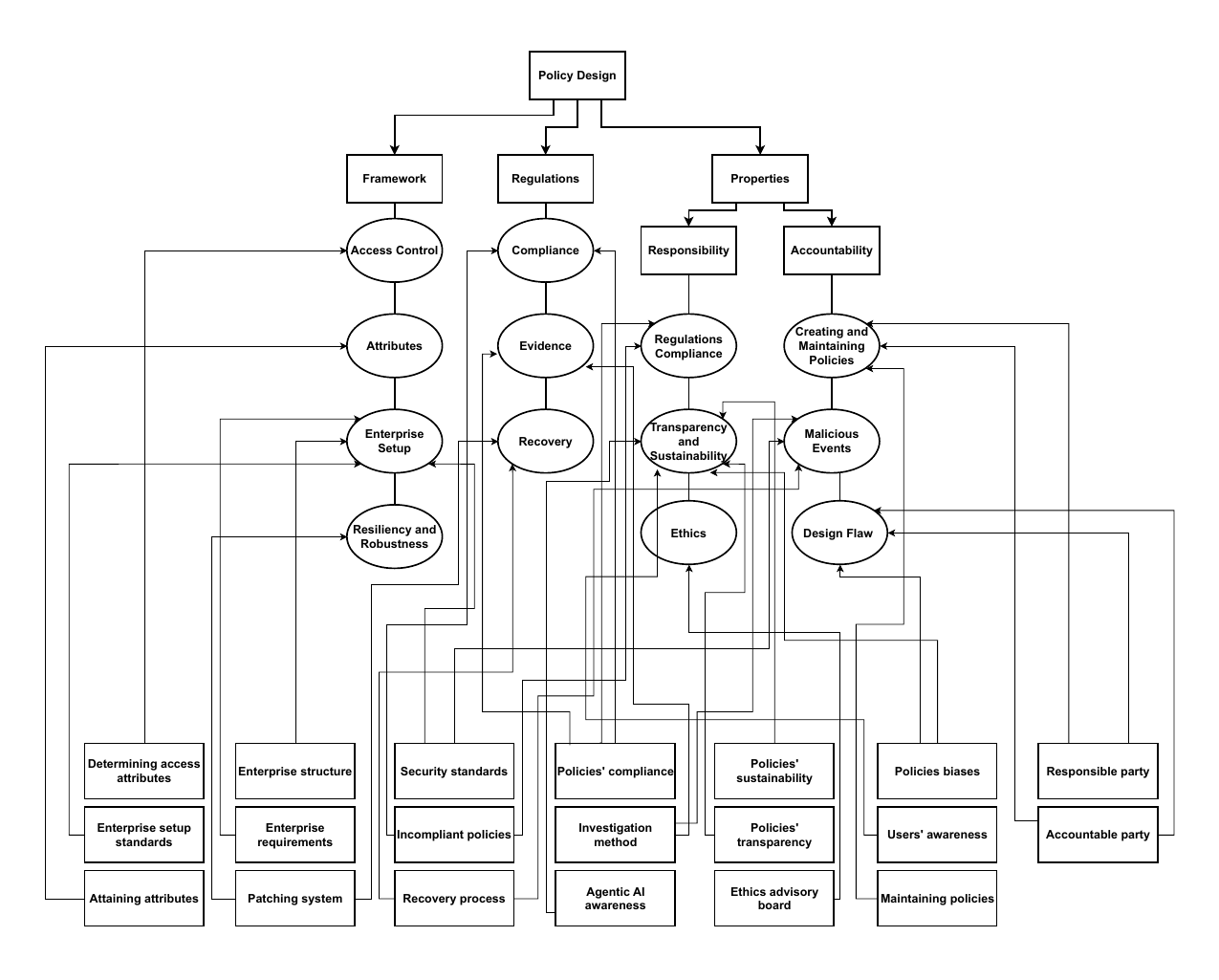}}
\caption{An illustration of challenges of policy design in ZTDN.}
\label{figure3}
\end{figure*}

Today, many policies are AI-generated; therefore, ensuring these policies comply with country-specific regulations is necessary. Responsible policy means that all attributes inside the policy have been thought of for all related components and have been defined clearly, and methods of due diligence must be in place. Ensuring these policies comply with local and international regulations and standards is also essential \cite{rose2020}. This involves the transparency of policies and rules, which can be achieved by clearly presenting these in SLAs, as discussed earlier. In addition, the sustainability of these components is vital to the growing digital footprint in technology and understanding their impact in real-world situations. Figure \ref{figure3} summarizes critical points when designing policy for enterprises and raises critical questions to form SLAs. Considerable issues regarding AI-generated policies are transparency and the associated biases, and it is essential to ensure ethical policy-making when AI is used for this purpose.

Jawhar \textit{et al.} (2024) \cite{jawhar2024} have developed a module to generate policies by AI systems using Open AI API. Their module which contains requirements on international standards compliance and list of controls in the security framework results in a prompt that are being fed into Chat GPT-4. Their experiment resulted in effective generated access control policies that are dynamically dependent on organization infrastructure and adhere to the international standards of NIST 800-171 and ISO 27001/27002. However, they have mentioned that these generated policies have to be audited and monitored to ensure that organizations are still in control of their security operations. Another example have been discussed by Fu \textit{et al.} (2025) \cite{fu2025} where they have used AI algorithms to detect and implement policies in order to produce better policies for the system in real time.

\begin{figure*}[h]
\centerline{\includegraphics[width=30pc]{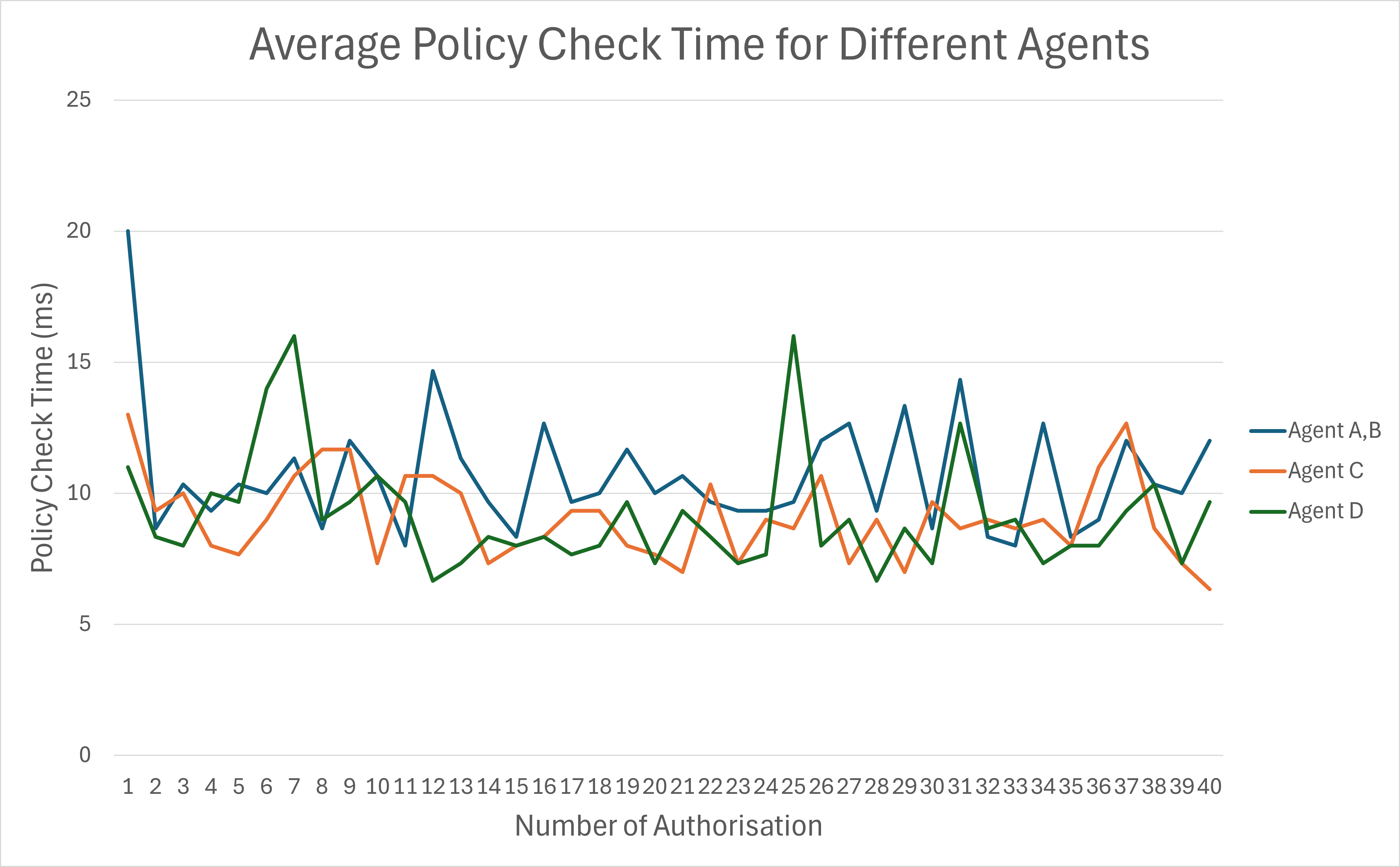}}
\caption{Average policy check time for three different agents doing different tasks.}
\label{figure4x}
\end{figure*}

\section{Fitting Zero-Trust (ZT) Into Current Standards}
ZT, in its application, may complement the already implemented cyber security standards in organizations. Current standards in cyber security include ISO 27001, which covers guidance for establishing, implementing, maintaining and continually improving an information security management system \cite{jahankhani2020}; NIST CSF ID.GV-1, which covers the establishment and communication of organizational cybersecurity policies and NIST Special Publication 800-53 contains a guide on Information Security Testing and Assessment \cite{nist2021}; and ISA/IEC 62443 that covers cybersecurity requirements and processes for implementing and maintaining electronically secure industrial automation and control systems (IACS) \cite{isa}. ZT standard for enterprises' cyber security itself is highlighted in NIST Special Publication 800-207, which explains in depth the definition of ZTA for enterprises and the different types of implementation \cite{rose2020}. 

Taking an example of relating ZT to ISO 27001 standard, the book by Jahankhani et al. (2020) \cite{jahankhani2020} have mentioned that ZT concepts do not cover physical security, culture, and governance, which are the core aspects of ISO 27001. Hence, ZT must be viewed as an augmentation to these standards, not a replacement. Adopting ZT can strengthen access control and network security control, which are key principles of ISO 27001. The standard requires regular review and improvement of the information security management system (ISMS). This aligns with the ZT principle of continuous monitoring. With ZT, it is always assumed that attackers are already inside the system, which aligns with a control category that ISO 27001 have: information security incident management (ISIM). Key ZT principles like verifying everything and implementing least privilege access can be beneficial in enhancing ISO 27001’s access control category. Additionally, ISO 27001 requires information segregation, which can be done by micro-segmentation, an aspect of ZT. This demonstrates effective access control implementation, network security management, and system acquisition, development, and maintenance, which shows the organization's risk management. 

\section{Securing Agentic AI With Zero-Trust (ZT)}
The increasing usage of the buzzword agentic AI has been prevalent in the technology industry lately. However, what agentic AI is and how it can help impact the security of the system are still critical questions to answer. Agentic AI is a term in which AI act as agents, replacing human administrators in performing tasks and automating processes. This can improve the efficiency of the system while at the same time reducing operational costs. Placing AI agents in the system can also help in making decisions faster. However, some issues regarding utilizing agentic AI in a system must be considered, which include security, privacy, and trust of the resources, and in the long-term operations, the issues would include traceability and auditability of agents.

There are quite a few security and privacy issues that have yet to be explored in terms of agentic AI. Firstly, there is the possibility of private information inside the resources being exposed by the agents. This also ties in with the second problem, which is how AI queries and results can be manipulated. For example, a malicious attacker can access AI agents to bypass the security system and obtain sensitive data and/or manipulate the system itself. This also leads to the third point, where AI agents can increase the likelihood of an attack by expanding the attack surface. For example, if an agent is compromised, it can be used as a gateway to access and exploit private information. Lastly, from the user's perspective, there is a social assumption that users can be impacted if AI handles sensitive data. AI requires access to data, as it needs it to train on to produce intelligent results. This can diminish users' confidence as they may not feel secure with the future developments of agentic AI and its role in distributed systems.
 
The idea of securing agentic AI with ZT can be explored to address these concerns. If an agent is placed as a point of contact for users before being able to access resources, ZT can be implemented on top of that to ensure that the results of the agent regarding access requests align with the policies that the PE has in place. This means ZT and its principles can benefit the system when incorporated into agentic AI systems. From the CAPEX/OPEX point of view, latency problems might arise in this case when the server for AI agents is overwhelmed with the number of request queries, which can lead to potential DoS attacks.

To further understand the implications of policy verification on the activity of agents, a four-agents scenario was developed, where two agents, A and B, were assigned the tasks of keyword search and keyword counting, respectively, for the content available on web links. The platform was developed in C\# using Blazor UI (client side) and .NET (server side). We used the House of Companies \cite{houseofcomp} information to create this scenario, which enables API-driven accessibility. Two additional agents, C and D, were created, which looked for company names and the associated officers, respectively. These agents were orchestrated using ZT policies. ZT components are implemented before users can access resources. Policy check times are obtained from the amount of time taken for the system to verify users’ credentials before allowing a request to be made to the specific agents. For example, we have added a feature that requires users to select their role, which is either administrator or normal user. This role will be stored locally and referred to whenever users access resources. When a request is made, the PEP will redirect that request to the PDP components, consisting of the PA and PE, located on the server side, to log the request and verify users’ credentials. In this example, user’s role will be verified. If the user is an administrator, then users will be allowed to continue their activity on the system. Policy administration logs these access requests and decisions and forwards them to PEP to open/close the communication channel to resources. To obtain these metrics, another agent is created to measure the time it takes to verify users with policy on the server side, as well as the time it takes for the PEP to send and receive requests and decisions on the client side. The analysis includes executing 50 continuous access requests to each agent with different tasks three times.

Based on the responsiveness of the agents, their average policy time is shown in Figure \ref{figure4x}. The graph shows that the average time it takes to verify policy between agents is almost similar throughout the execution of the access request. Both agent A and agent B are working concurrently to execute the task where users input a link to a website. Hence, their average policy verification time is combined by checking the policies before both agents are executed to do the task. The average policy time also depends on the network and service accessibility, which are beyond the scope of this work, and may vary when performed in a different setting and infrastructure. Irrespective of this, the implementation proves that augmenting agentic AI with ZT provides an extra layer of security by constantly authenticating and authorizing users whenever a request is made without major performance tradeoffs.

\section{Formal Verification of Policies in ZTDN}

\begin{figure*}[!ht]
\centering
    \begin{subfigure}{.49\textwidth}
        \centering
        \includegraphics[width=.7\linewidth, trim=0 180 0 180, clip]{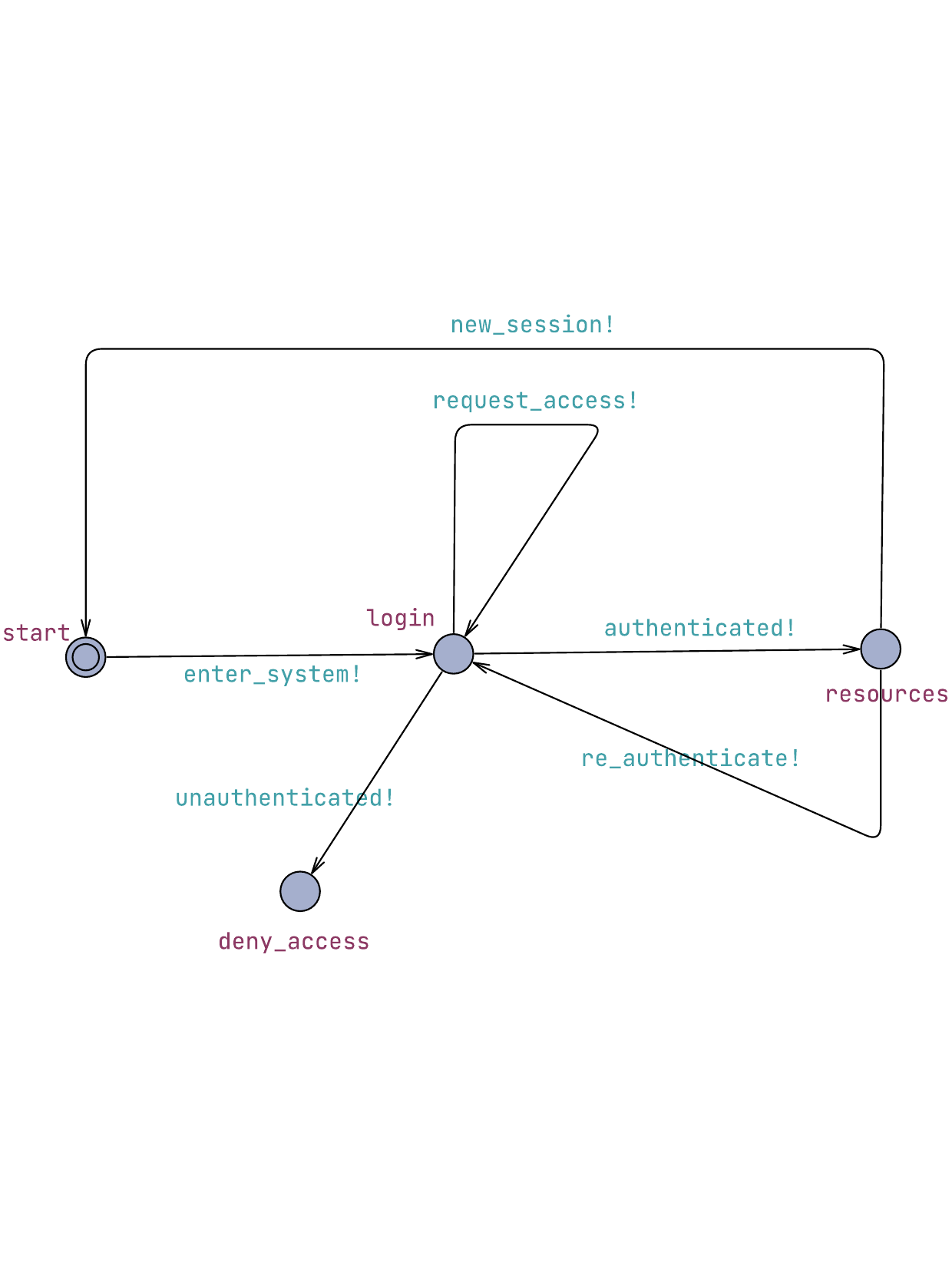}
        \caption{User}
        \label{fig:sfig1}
    \end{subfigure}
    \begin{subfigure}{.49\textwidth}
        \centering
        \includegraphics[width=.7\linewidth, trim=0 240 0 260, clip]{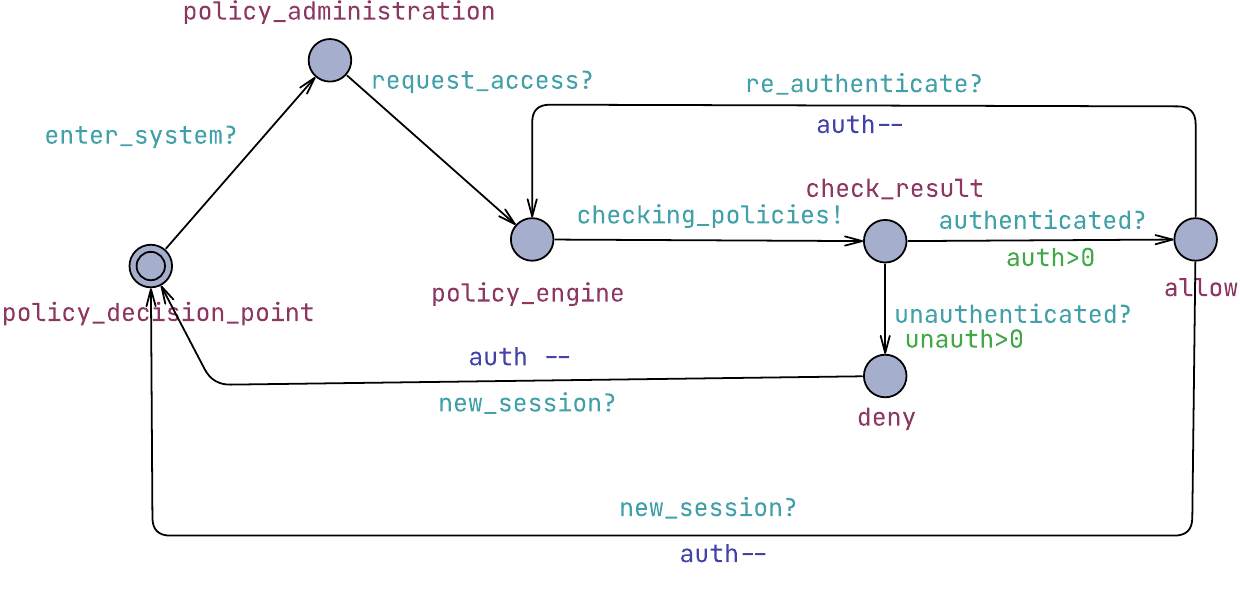}
        \caption{Policy decision point}
        \label{fig:sfig2}
    \end{subfigure}
    \begin{subfigure}{.49\textwidth}
        \centering
        \includegraphics[width=1\linewidth,  trim=0 280 0 280, clip]{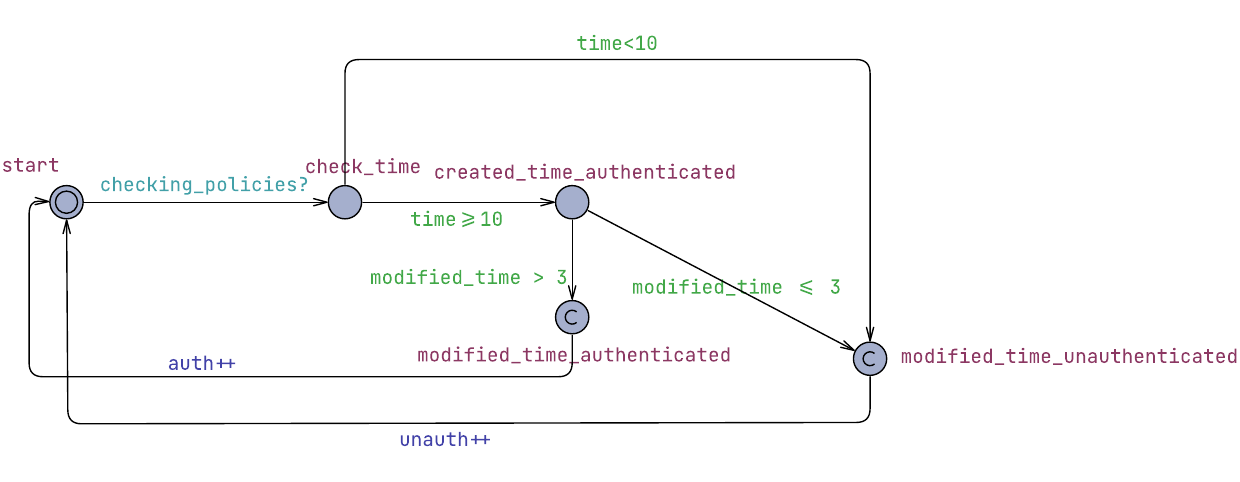}
        \caption{Policy engine}
        \label{fig:sfig3}
    \end{subfigure}
    \caption{An example of ZTDN's state diagram for a single user.}
    \label{figure4}
\end{figure*}

\begin{figure*}[!ht]
\centerline{\includegraphics[width=30pc]{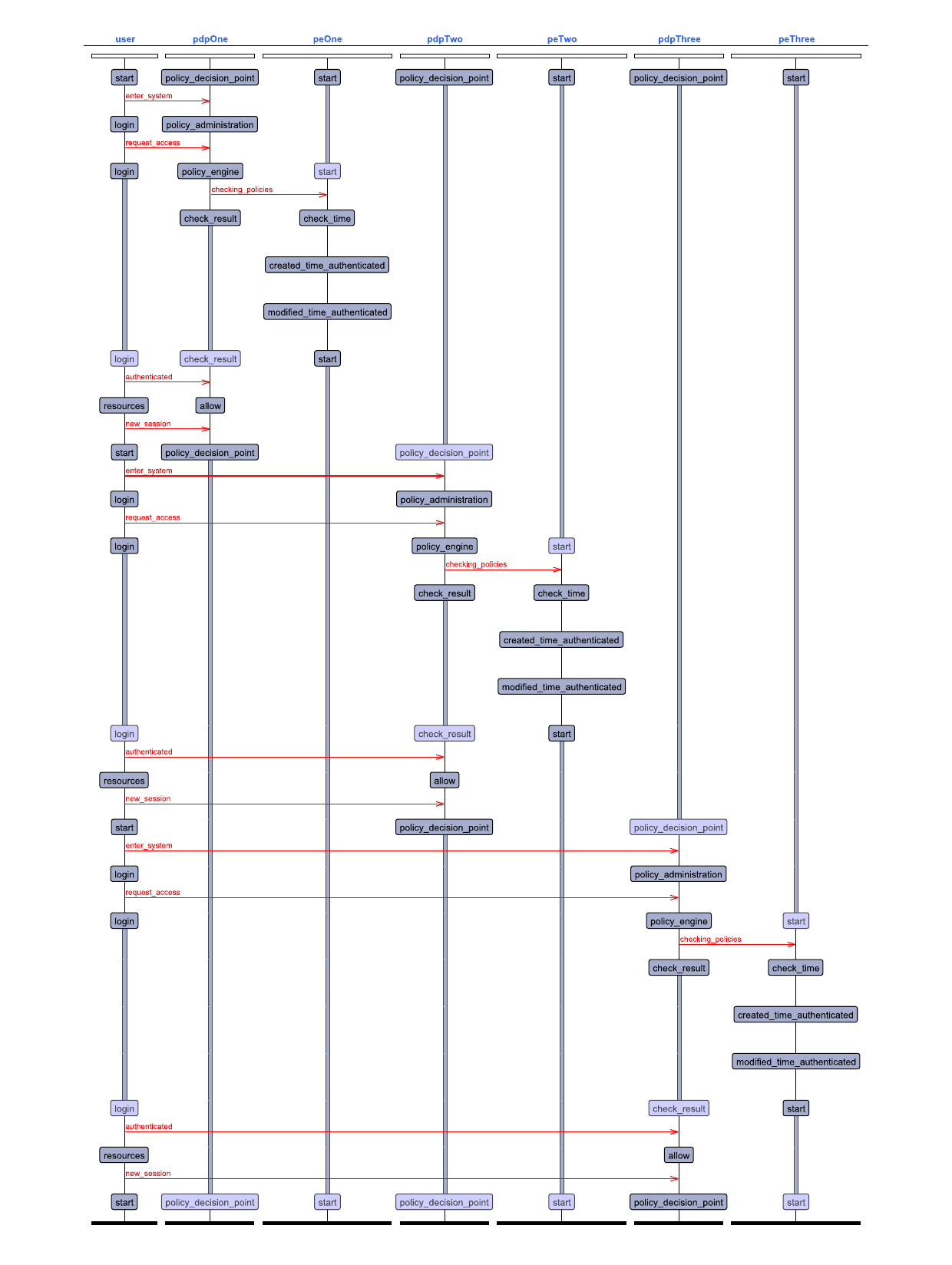}}
\caption{ZTDN's state diagram showing the authenticated request flow.}
\label{figure5}
\end{figure*}

With the number of policies that enterprises have for input to the PE, as shown in Figure \ref{figure1}, formal verification of those policies is crucial. This ensures that there are no errors in the policies that might affect the working of the system while also reducing the probability of malicious attacks in the system. However, with the rapid increase of usage of AI in this space, it does get more complicated to maintain policies in organizations, along with the dynamic adaptive policies, which is a part of ZT's tenet. This can be further impacted with generative AI (GenAI). Hence, formally verifying these dynamic policies is beneficial for enterprise systems. This is supported by the research done by Kanner and Kanner (2021) \cite{kanner2021}, where it has been mentioned that verification of policies, especially access control policies, is a helpful step in the development of the system to guard unauthorized access. Verification also includes formally establishing an explanation for every single case of implementation in the system. The authors \cite{kanner2021} have explained the pros of verifying policies, which include automatically finding errors in the design/implementation of the system.

Implementing micro-segmentation is a way to apply ZT in enterprise systems \cite{keeriyattil2019}. With different locations hosting networks for the enterprise, it becomes natural to implement micro-segmentation, and AI-generated SLAs can drive the decision to implement this micro-segmentation. This can help prevent all servers from being maliciously attacked at the same time. For example, by having all enterprise-distributed networks micro-segmented, an attack on a component can be compartmentalized in a specific area without spreading laterally through the system. This also aligns with the service migration in the enterprise network, as compartmentalization can benefit them by ensuring the system's security through micro-segmentation of its components and allowing for gradual service migration. 

A method that can be used to do formal verifications of policies in ZTDN can be attained using timed automata (TA) \cite{alur1994}. Using this theory, it can trace back the time when policies are made, as well as if there is any point after that time when the policies have been changed. This can be helpful to verify policies that are fed into the PE, and ensure these are legitimately valid and have not been tampered with by any party. This is particularly useful if AI generates some of the policies, and time stamping is required against them.

It has not been widely explored whether policies created using AI-powered tools are deemed trustworthy. Further, aligning this concept with ZT, it is essential that at every step in the network, no access is trusted and must be authenticated and authorized. To complement the use of TA, tools like, UPPAAL \cite{bengtsson1995} can be helpful. UPPAAL is an open-source tool that act as a protocol verifier and can be applied to verify and validate policies. In this paper, a case-study of formally verifying policies using UPPAAL is explored in the context of ZTDN.

Figure \ref{figure4} illustrates an example of UPPAAL state diagrams from the user, PDP, and PE perspectives related to the distributed scenario in Figure \ref{figure2}. Figure \ref{fig:sfig1} shows the system for users, which shows the process to request access to resources, and it synchronizes with the processes in the PDP. Figure \ref{fig:sfig2} depicts the process that takes place in the PDP component of a ZTDN. This shows that the PDP will always re-authenticate users even after being authenticated to access resources the first time. 

The Figure \ref{fig:sfig3} illustrates how a policy can be verified in the PE. If the policy has been altered recently without any traceability component and missing log, there is reason to believe that malicious activity has occurred. Users will not be able to access resources as a result. When access is requested, there will also be checks to determine whether policies have been modified since the request has been received. This can enhance the PE's level of security. Simulating these processes would make it possible to detect occasions in which policies have been violated or altered by malicious actors or compromised AI agents.

To further demonstrate the flow of the PE in the PDP component of a ZTDN, Figure \ref{figure5} shows the detail of the processes of each state in each template and how all of the components are interconnected. It also shows that there are no deadlock processes, as the re-authentication process of the user after getting access to the resources is always available and ready to be executed. This Figure \ref{figure5} is particularly relevant to the scenario in Figure \ref{figure2} where a user attempts to gain entry to three different enterprise networks.

\section{Policy Decision and Enforcement in ZTDN}

\begin{table*}[!ht]
    \caption{Policy decision and enforcement in ZTDN.}
    \centering
    \renewcommand{\arraystretch}{1.2}
    \begin{tabular}{|p{13mm}|p{17mm}|p{22mm}|c|c|c|c|}
    \hline
        \centering Focus & \centering Properties & \centering Metrics and Influences & \multicolumn{1}{c|}{Adversarial Risks} & \multicolumn{1}{c|}{Entities} & \multicolumn{1}{c|}{Instances} & \multicolumn{1}{c|}{Accountability} \\ 
    \hline
        \multirow{8}{13mm}{ \centering\textbf{Security}} & Availability & Service and policy check time (internal) & \multirow{15}{*}{\rotatebox[origin=c]{90}{\parbox{100mm} { \centering Entity misbehavior\newline Policy misconfiguration\newline  Policy engine exploitation\newline Insider threats\newline Multi-factor authentication failure \newline}}} &
  
        \multirow{15}{*}{\rotatebox[origin=c]{90}{\parbox{100mm} { \centering End users\newline Policy decision points\newline Policy enforcement points\newline Payload in disaggregated Scenarios \newline}}} &
  
        \multirow{15}{*}{\rotatebox[origin=c]{90}{\parbox{100mm} { \centering Better load sharing\newline Trust within zero-trust\newline Application-specific policing\newline Adaptive and flexible properties\newline Robust integrations\newline}}} &
  
        \multirow{15}{*}{\rotatebox[origin=c]{90}{\parbox{100mm} { \centering Policy vs Third-party infrastructure\newline Enterprise network\newline \centering Compliance and regulations\newline Agentic AI forensics \newline}}} \\ \cline{2-3} &
  
        Integrity & \RaggedRight Data checksums and policy mapping (internal) & & & & \\ \cline{2-3} &
  
        Confidentiality & Access attempts (external) & & & & \\ \cline{2-3} &
        Authorization & Access violations (external) & & & & \\ \cline{2-3} &
        Auditability & Logs and compliance (internal) & & & & \\ \cline{2-3} &
        Traceability & \RaggedRight Entity behavior logs (internal) & & & & \\ \cline{2-3} &
        Authentication & \RaggedRight Identity Verification (internal) & & & & \\ \cline{2-3} &
        Robustness & Entity compromise (internal) & & & & \\ \cline{1-3}

        \multirow{7}{13mm}{ \centering\textbf{Functional Safety}} &  Tolerance & Faults (internal and external) & & & & \\ \cline{2-3} &
        Dependencies & Service availability (external) & & & & \\ \cline{2-3} &
        Diagnosis & Service availability (external) & & & & \\ \cline{2-3} &
        Capacity & \RaggedRight Service rate and planning (external) & & & & \\ \cline{2-3} &
        Migration & Service availability (external) & & & & \\ \cline{2-3} &
        Containment & \RaggedRight Service availability and compartmentalization (internal and external) & & & & \\ \cline{2-3} &
        Recovery & Faults and failures (internal and external) & & & & \\ \hline
    \end{tabular}
    \label{table_2}
\end{table*}

When discussing policy decisions and enforcement in ZTDN, security and functional safety are both the focus of the technology in ensuring straightforward interaction between users and the service. There are a few properties in each focus that should be monitored, followed by the metrics by which each property is measured, as shown in Table~\ref{table_2}. These include availability, which can be observed by service and policy check time; integrity by doing data checksums and policy mapping; confidentiality by measuring the number of access attempts to service; authorization, which can be considered from the number of access violations; auditability viewed by logs and compliance; traceability which is considered by entity behaviour logs; authentication by doing identity verification; and robustness by measuring the amount of entity compromise.

On the other side, there are also a few functional safety properties that should be considered in terms of making decisions and enforcing policies in ZTDN. These include tolerance, which can be seen by faults observation; dependencies and diagnosis, which both can be observed by service availability; capacity, shown by the safety rate and planning; migration, which can be observed by service availability; containment, which can be observed by both service availability and compartmentalization; and recovery which the number of faults and failures of service can be measured.

Risks will always be present for any service, and in terms of the security and functional safety of policy decisions and enforcement in ZTDN, there are a few notable risks. Hence, there needs to be cautionary measures to ensure that the service is constantly monitored to check for any suspicious activity. These efforts are all vital in guarding the services against adversaries such as entity misbehaviour, policy misconfiguration, PE exploitation, insider threats, and multi-factor authentication failure. Another crucial part is accountability within policies in ZTDN, which can be managed by looking into policy vs. third-party infrastructure, enterprise network, compliance and regulations, and agentic AI forensics.

\section{Open Problems and Research Directions}

There are several open questions related to the expansion of ZTDN and its integration with agentic AI, in particular, where the focus is on having a responsible policy design.  One of the critical directions that remains open is the run-time verification of the policies that are used as inputs for the PE component.  Niu \textit{et al.} (2022) \cite{niu2022} have proposed a ZT security policy detection method based on online verification to evaluate the effectiveness and security of policies in the PA component.  It is worth researching other methods of formal verification for policies by considering the trade-off between the complexity of verification and the number of properties checked.  Other methods of policy verification can help to improve the security and efficiency of policies. Using UPPAAL as a tool is just one way to perform formal verification of policies, and there are other methods that can be employed to ensure policy compliance and prevent malicious events \cite{niu2022}, \cite{liu2025}.  Another direction to explore is having a clearly defined methodology specifically catered to ZTDN that can be implemented industry-wide. An example of this is provided by Li \textit{et al.} (2025) \cite{li2025}, who used ZT as a verifier to address security issues found in Compute First Networks (CFN) to be used as a platform for AI-generated content (AIGC) services.  However, there is still a gap in the literature when it comes to implementing ZT on agentic AI systems. Another less explored yet critical dimension of policy design is establishing a clear structure for those overseeing decisions on creating, monitoring, and removing policies in ZTDN, ensuring stability and accountability. This is supported by research conducted by Køien (2024) \cite{koien2024}, which stresses the urgent need to assess transparency and accountability in the system, especially when AI is involved. It is vital to ensure ethical policy-making when AI-driven policies are involved. This includes associated biases that come with generating policies using AI. The above open problems are summarized as follows:
\begin{itemize}
    \item \textbf{AI-driven policies}: The ethical side of having AI-driven policies needs to be further investigated, including their impact on the systems. Here, the impact can be positive or negative, and it will be driven by the set of requirements used for deriving the policies~\cite{fu2025}, \cite{sharma2025}. Further, AI-driven policies can help run scenarios before any changes are reflected in the system.
    \item \textbf{Responsibility and accountability for policy design}: This is a large area to explore when dealing with policy design in ZTDN.  There needs to be clear legislation on the responsibility and accountability of creating, designing, and implementing policies, especially when AI is involved and may present security risks. Some points to explore include having a traceability system for policies \cite{koien2024}, \cite{adom2024}. The use of agentic protocols, along with telemetry information gathering, could help build solutions that are traceable and offer accountability over policies. 
    \item \textbf{Agent orchestrator for ZTDN}: Agentic AI leverages an orchestrator for managing tasks and associated agents. However, when operating in ZTDN, the roles of PE and orchestrator can be combined for better visibility and control over agents, which may help reduce the complexity of having two separate entities that can be unified for better CAPEX/OPEX. Further exploration of the actual implementation of this can be conducted by examining the findings on the effectiveness and security of agentic AI systems when combined with ZT \cite{li2025}, \cite{haque2024}. 
    \item \textbf{Standards for policy design in ZTDN}: A major aspect of ZT is the standards that are being used currently in this field. As previously mentioned, current standards have not yet accounted for state-of-the-art technology advancements. Hence, there needs to be discussions on updating these standards to tailor them towards security risks in policy design \cite{rose2020}, \cite{al2024}. A clearly defined standard for attributes that takes into account policy design elements, such as trust scores or levels, thresholds, and when a user is allowed access to systems, is another associated area to address. Standards \cite{rose2020} that are available do not account for factors such as agentic AI, and with recent AI advancements, guidelines for AI forensics could be a major open issue.
\end{itemize}

\section{Conclusion}
This article has discussed the challenges and solutions of policy design in zero-trust distributed networks (ZTDN). Details of where ZTDN fits into the current standards and possible solutions for verifying policies are also discussed. The article further presents aspects of agentic AI and the enforcement and decision-making process by including results on policy verification time for agents. It provides further details about the formal verification of policies in ZTDN, which is shown by a case study using timed automata through UPPAAL. This shows the feasibility of formally verifying and validating policies, emphasizing the policies have not been modified or tampered with. Future work on leveraging agentic AI for this purpose can be done along with the important aspects of accountability, traceability and forensics.


\bibliographystyle{ieeetr}
\bibliography{references.bib}

\end{document}